\documentclass[aps,pra,twocolumn,noeprint,superscriptaddress,amsmath,amssymb]{revtex4-2}

\usepackage{graphicx}
\usepackage{dcolumn}
\usepackage{amsmath}
\usepackage{csquotes}
\usepackage{braket}
\usepackage{bm}
\usepackage{multirow}
\usepackage{comment}
\usepackage{float}
\usepackage[dvipsnames]{xcolor}
\usepackage{hyperref}
\hypersetup{
    colorlinks=true,
    citecolor=Blue,
    linkcolor=Blue,
    urlcolor=Blue
}
\usepackage{booktabs}
\usepackage{siunitx}
\DeclareSIUnit\gauss{G}
\DeclareSIUnit\electron{\textit{e}}

\graphicspath{ {./images/} }

\begin{document}

\title{A new bound on the electron's electric dipole moment}

\author{Tanya S. Roussy}
\thanks{These authors contributed equally to this work.}
\affiliation{JILA, NIST and University of Colorado, Boulder, Colorado 80309, USA}
\affiliation{ Department of Physics, University of Colorado, Boulder, Colorado 80309, USA}

\author{Luke Caldwell}
\thanks{These authors contributed equally to this work.}
\affiliation{JILA, NIST and University of Colorado, Boulder, Colorado 80309, USA}
\affiliation{ Department of Physics, University of Colorado, Boulder, Colorado 80309, USA}

\author{Trevor Wright}
\affiliation{JILA, NIST and University of Colorado, Boulder, Colorado 80309, USA}
\affiliation{ Department of Physics, University of Colorado, Boulder, Colorado 80309, USA}

\author{William B. Cairncross}
\altaffiliation[Present address: ]{Atom Computing, 918 Parker Street, suite A-13, Berkeley California, 94710}
\affiliation{JILA, NIST and University of Colorado, Boulder, Colorado 80309, USA}
\affiliation{ Department of Physics, University of Colorado, Boulder, Colorado 80309, USA}

\author{Yuval Shagam}
\altaffiliation[Present address: ]{Schulich Faculty of Chemistry, Technion - Israel Institute of Technology, Haifa 3200003, Israel}
\affiliation{JILA, NIST and University of Colorado, Boulder, Colorado 80309, USA}
\affiliation{ Department of Physics, University of Colorado, Boulder, Colorado 80309, USA}

\author{Kia Boon Ng}
\affiliation{JILA, NIST and University of Colorado, Boulder, Colorado 80309, USA}
\affiliation{ Department of Physics, University of Colorado, Boulder, Colorado 80309, USA}

\author{Noah Schlossberger}
\affiliation{JILA, NIST and University of Colorado, Boulder, Colorado 80309, USA}
\affiliation{ Department of Physics, University of Colorado, Boulder, Colorado 80309, USA}

\author{Sun Yool Park}
\affiliation{JILA, NIST and University of Colorado, Boulder, Colorado 80309, USA}
\affiliation{ Department of Physics, University of Colorado, Boulder, Colorado 80309, USA}

\author{Anzhou Wang}
\affiliation{JILA, NIST and University of Colorado, Boulder, Colorado 80309, USA}
\affiliation{ Department of Physics, University of Colorado, Boulder, Colorado 80309, USA}

\author{Jun Ye}
\affiliation{JILA, NIST and University of Colorado, Boulder, Colorado 80309, USA}
\affiliation{ Department of Physics, University of Colorado, Boulder, Colorado 80309, USA}

\author{Eric A. Cornell}
\affiliation{JILA, NIST and University of Colorado, Boulder, Colorado 80309, USA}
\affiliation{ Department of Physics, University of Colorado, Boulder, Colorado 80309, USA}


\begin{abstract}%
The Standard Model cannot explain the dominance of matter over anti-matter in our universe \cite{Dine2003}. This imbalance indicates undiscovered physics that violates combined CP symmetry \cite{Sakharov1967,Gavela1994}. Many extensions to the Standard Model seek to explain the imbalance by predicting the existence of new particles \cite{Engel2013}. Vacuum fluctuations of the fields associated with these new particles can interact with known particles and make small modifications to their properties; for example, particles which violate CP symmetry will induce an electric dipole moment of the electron (eEDM). The size of the induced eEDM is dependent on the masses of the new particles and their coupling to the Standard Model. To date, no eEDM has been detected, but increasingly precise measurements probe new physics with higher masses and weaker couplings. Here we present the most precise measurement yet of the eEDM using electrons confined inside molecular ions, subjected to a huge intra-molecular electric field, and evolving coherently for up to \SI{3}{\second}. Our result is consistent with zero and improves on the previous best upper bound \cite{ACME2018} by a factor  $\sim2.4$. Our sensitivity to $10^{-19}$\,eV shifts in molecular ions provides constraints on broad classes of new physics above $10^{13}$\,eV, well beyond the direct reach of the LHC or any other near- or medium-term particle collider.
\end{abstract}
 

\newcommand{\Cdoub}{{\mathcal C}_{\rm d}}

\newcommand{\numblocks}{1329}

\newcommand{\tPz}{{}^3\Pi_{0^+}}
\newcommand{\tPzm}{{}^3\Pi_{0^-}}
\newcommand{\sSp}{{}^1\Sigma^+}
\newcommand{\tDo}{{}^3\Delta_1}
\newcommand{\tSm}{{}^3\Sigma_{0^+}^-}

\newcommand{\ltrans}{$\mathcal{L}_{\rm trans}^{961}$}
\newcommand{\lvc}{$\mathcal{L}_{\rm vc}^{818}$}
\newcommand{\lop}{$\mathcal{L}_{\rm op}^{1082}$}
\newcommand{\ldepl}{$\mathcal{L}_{\rm depl}^{814}$}

\newcommand{\gFul}{g_F^{u/l}}
\newcommand{\gFbar}{\bar{g}_F}
\newcommand{\dgF}{{\delta g_{F}}}
\newcommand{\dgeff}{{\delta g}_{\rm eff}}
\newcommand{\dmf}{d_{{\rm mf}}}
\newcommand{\wef}{\omega_{ef}}
\newcommand{\fef}{f_{ef}}
\newcommand{\Ehf}{E_{\rm hf}}
\newcommand{\Apar}{A_\parallel}
\newcommand{\Gpar}{G_\parallel}
\newcommand{\Eeff}{{\mathcal E}_{\rm eff}}
\newcommand{\dDelta}{{\delta \Delta}}
\newcommand{\Deltabar}{\bar{\Delta}}

\newcommand{\fn}{f^0}
\newcommand{\fB}{f^B}
\newcommand{\fD}{f^D}
\newcommand{\fR}{f^R}
\newcommand{\fBD}{f^{DB}}
\newcommand{\fBR}{f^{BR}}
\newcommand{\fDR}{f^{DR}}
\newcommand{\fBDR}{f^{DBR}}
\newcommand{\fI}{f^I}
\newcommand{\fBI}{f^{BI}}
\newcommand{\fDI}{f^{DI}}
\newcommand{\fRI}{f^{RI}}
\newcommand{\fDBI}{f^{DBI}}
\newcommand{\fBRI}{f^{BRI}}
\newcommand{\fDRI}{f^{DRI}}
\newcommand{\fBDRI}{f^{DBRI}}

\newcommand{\Erot}{{\mathcal E}_{\rm rot}}
\newcommand{\hatErot}{\hat{\mathcal E}_{\rm rot}}
\newcommand{\Vrot}{V_{\rm rot}}
\newcommand{\Vrf}{V_{\rm rf}}
\newcommand{\vecErot}{\vec{\mathcal E}_{\rm rot}}
\newcommand{\rrot}{r_{\rm rot}}

\newcommand{\Brot}{{\mathcal B}_{\rm rot}}
\newcommand{\vecBrot}{\bm{\mathcal B}_{\rm rot}}
\newcommand{\Bperp}{{\mathcal B}_{\perp}}
\newcommand{\vecBperp}{\bm{\mathcal B}_{\perp}}
\newcommand{\BX}{{\mathcal B}_X}
\newcommand{\BY}{{\mathcal B}_Y}
\newcommand{\BZ}{{\mathcal B}_Z}
\newcommand{\Bx}{{\mathcal B}_x}
\newcommand{\By}{{\mathcal B}_y}
\newcommand{\Bz}{{\mathcal B}_z}
\newcommand{\EX}{{\mathcal E}_X}
\newcommand{\EY}{{\mathcal E}_Y}
\newcommand{\EZ}{{\mathcal E}_Z}
\newcommand{\Ex}{{\mathcal E}_x}
\newcommand{\Ey}{{\mathcal E}_y}
\newcommand{\Ez}{{\mathcal E}_z}
\newcommand{\Eperp}{{\mathcal E}_\perp}
\newcommand{\Baxgrad}{{\mathcal B}'_{\rm axgrad}}
\newcommand{\Btrans}{{\mathcal B}'_{\rm trans}}
\newcommand{\Bone}{{\mathcal B}'_{1}}
\newcommand{\Btwo}{{\mathcal B}'_{2}}
\newcommand{\Bthree}{{\mathcal B}'_{3}}
\newcommand{\Baxgradnr}{{\mathcal B}_{\rm axgrad}^{\prime{\rm nr}}}
\newcommand{\Btransnr}{{\mathcal B}_{\rm trans}^{\prime{\rm nr}}}
\newcommand{\Bonenr}{{\mathcal B}_{1}^{\prime{\rm nr}}}
\newcommand{\Btwonr}{{\mathcal B}_{2}^{\prime{\rm nr}}}
\newcommand{\Bthreenr}{{\mathcal B}_{3}^{\prime{\rm nr}}}
\newcommand{\Brotnr}{{\mathcal B}_{\rm rot}^{\rm nr}}
\newcommand{\Bxnr}{{\mathcal B}_x^{\rm nr}}
\newcommand{\Bynr}{{\mathcal B}_y^{\rm nr}}
\newcommand{\Bznr}{{\mathcal B}_z^{\rm nr}}
\newcommand{\BXnr}{{\mathcal B}_X^{\rm nr}}
\newcommand{\BYnr}{{\mathcal B}_Y^{\rm nr}}
\newcommand{\BZnr}{{\mathcal B}_Z^{\rm nr}}

\newcommand{\wrot}{\omega_{\rm rot}}
\newcommand{\vecwrot}{\bm{\omega}_{\rm rot}}
\newcommand{\wrf}{\omega_{\rm RF}}
\newcommand{\frot}{f_{\rm rot}}
\newcommand{\Trot}{T_{\rm rot}}
\newcommand{\vecfrot}{\bm{f}_{\rm rot}}
\newcommand{\frf}{f_{\rm rf}}
\newcommand{\wsec}{\omega_{\rm sec}}

\newcommand{\Bm}{{\mathcal B}}
\newcommand{\Em}{{\mathcal E}}
\newcommand{\hffp}{HfF$^+$}
\newcommand{\hfp}{Hf$^+$}
\newcommand{\td}{{^3\Delta_1}}
\newcommand{\orderof}{{\mathcal O}}
\newcommand{\sgn}{{\rm sgn}}
\newcommand{\stat}{{\rm stat}}
\newcommand{\syst}{{\rm syst}}
\newcommand{\mus}{\,\mu{\rm s}}
\newcommand{\percm}{\,{\rm cm}^{-1}}
\newcommand{\mVcm}{\,{\rm mV/cm}}
\newcommand{\Vcm}{\,{\rm V/cm}}
\newcommand{\Volt}{\, {\rm V}}
\newcommand{\musmm}{\,\mu{\rm s/mm}}
\newcommand{\Kelvin}{\,{\rm K}}

\newcommand{\tilB}{{\tilde{B}}}
\newcommand{\tilD}{{\tilde{D}}}
\newcommand{\tilR}{{\tilde{R}}}
\newcommand{\tilI}{{\tilde{I}}}
\newcommand{\tilP}{{\tilde{P}}}
\newcommand{\tilBD}{\widetilde{BD}}
\newcommand{\tilBR}{\widetilde{BR}}
\newcommand{\tilDR}{\widetilde{DR}}
\newcommand{\tilBDR}{\widetilde{BDR}}
\newcommand{\tilS}{{\tilde{S}}}
\newcommand{\tils}{{\tilde{s}}}

\newcommand{\Heff}{H_{\rm eff}}
\newcommand{\Htum}{H_{\rm tum}}
\newcommand{\Hhf}{H_{\rm hf}}
\newcommand{\HS}{H_{\rm S}}
\newcommand{\HZe}{H_{{\rm Z},e}}
\newcommand{\HZN}{H_{{\rm Z},N}}
\newcommand{\Hrot}{H_{\rm rot}}
\newcommand{\HOD}{H_{\Omega}}
\newcommand{\Hedm}{H_{\rm edm}}


\newcommand{\pdet}{p_{\rm det}}

\newcommand{\Expect}[1]{{\rm E}\left( #1 \right)}
\newcommand{\Var}[1]{{\rm Var}\left( #1 \right)}
\newcommand{\Std}[1]{{\rm Std}\left( #1 \right)}
\newcommand{\Cov}[1]{{\rm Cov}\left( #1 \right)}
\newcommand{\Kurt}[1]{{\rm Kurt}\left( #1 \right)}
\newcommand{\threej}[6]{\begin{pmatrix} 
    #1 & #3 & #5 \\ 
    #2 & #4 & #6 
    \end{pmatrix}}
\newcommand{\sixj}[6]{\begin{Bmatrix} 
    #1 & #2 & #3 \\ 
    #4 & #5 & #6 
    \end{Bmatrix}}
\newcommand{\ME}[3]{\left\langle #1 \left| #2  \right| #3 \right\rangle}
\newcommand{\RME}[3]{\left\langle #1 \left\| #2 \right\| #3 \right\rangle}
\newcommand{\xmark}{$\bm{\times}$}

\newcommand{\wbc}[1]{\textcolor{red!50!black}{$^{\textrm{wbc}}${#1}}}
\newcommand{\wbcx}[1]{\todo[author=wbc,inline,color=red!30]{#1}}

\newcommand{\fcorrA}{-1}
\newcommand{\sfcorrA}{5}
\newcommand{\sftotA}{5}
\newcommand{\fcorrB}{-3}
\newcommand{\sfcorrB}{2}
\newcommand{\sftotB}{4}
\newcommand{\fcorrC}{-1}
\newcommand{\sfcorrC}{1}
\newcommand{\sftotC}{2}
\newcommand{\fcorrD}{0}
\newcommand{\sfcorrD}{0}
\newcommand{\sftotD}{0}
\newcommand{\fcorrE}{6}
\newcommand{\sfcorrE}{13}
\newcommand{\sftotE}{14}
\newcommand{\fcorrF}{-2}
\newcommand{\sfcorrF}{8}
\newcommand{\sftotF}{8}
\newcommand{\fcorrG}{-107}
\newcommand{\sfcorrG}{163}
\newcommand{\sftotG}{195}
\newcommand{\feedmmHz}{0.10}
\newcommand{\sstatmHz}{0.87}
\newcommand{\ssystmHz}{0.20}
\newcommand{\sstatuHz}{868}
\newcommand{\ssystuHz}{195}
\newcommand{\stotuHz}{890}
\newcommand{\deecm}{0.9}
\newcommand{\sstatecm}{7.7}
\newcommand{\ssystecm}{1.7}
\newcommand{\ubecm}{1.3}
\newcommand{\mhztoecm}{1.13}

\newcommand{\vecpi}[0]{{\bm \pi}}
\newcommand{\vecPi}[0]{{\bm \Pi}}
\newcommand{\vecsigma}[0]{{\bm \sigma}}
\newcommand{\veca}[0]{{\bf a}}
\newcommand{\vecA}[0]{{\bf A}}
\newcommand{\vecx}[0]{{\bf x}}
\newcommand{\vecB}[0]{{\bf B}}
\newcommand{\vecalpha}[0]{{\bm \alpha}}
\newcommand{\vecomega}[0]{{\bm \omega}}
\newcommand{\vecL}[0]{{\bf L}}
\newcommand{\vecJ}[0]{{\bf J}}
\newcommand{\vecE}[0]{{\bm{\mathcal E}}}

\maketitle



Electric dipole moments of fundamental particles, like the electron, are signatures of time-reversal symmetry violation---equivalent to violation of combined charge and parity (CP) symmetry \cite{Khriplovich1997}. CP symmetry is broken in the Standard Model but only in the quark sector \cite{Patrignani2016}, so the coupling to leptons is weak and the predicted eEDM several orders of magnitude below current experimental sensitivity \cite{Yamaguchi2020, Ema2022}. Explaining the imbalance of matter and anti-matter in the universe requires additional CP violation, beyond that present in the Standard Model \cite{Gavela1994}. Many extensions have been proposed which add new particles at energies higher than any so far discovered, with CP-violating interactions. These new particles can induce a much larger eEDM, often within reach of near-term experiments \cite{Pospelov2005,Engel2013,Nakai2017}. A non-zero measurement at current experimental sensitivities would unambiguously signal new physics. Our measurement uses quantum-projection-noise limited spectroscopy on samples of hundreds of molecular ions with interrogation times of up to \SI{3}{\second}. Our result, $d_e = \SI[parse-numbers=false]{(-1.3 \pm 2.0_{\rm stat} \pm 0.6_{\rm syst})\times 10^{-30}}{\electron\centi\meter}$, is consistent with zero and gives an upper bound of $|d_e|<\SI{4.1e-30}{\electron\centi\meter}$ at 90\% confidence. 

An eEDM $\vec{d_e}=d_e\hat{s}$---with $\hat{s}$ a unit vector along the spin of the electron---subject to an electric field $\vec{\mathcal{E}}$ has an energy $-\vec{d_e}\cdot\vec{\mathcal{E}}$. The essence of an eEDM search is to measure the energy shift when $\hat{s}$ is aligned with $\vec{\mathcal{E}}$ compared to when it is anti-aligned. The size of the observable shift scales with the size of $\vec{\mathcal{E}}$ and thus many existing \cite{Hudson2011,Cairncross2017,ACME2018} and proposed \cite{Kozyryev2017,Vutha2018,Aggarwal2018,Fitch2020} eEDM experiments use electrons embedded inside polar molecules where intramolecular electric fields can be $\sim10^5$ times larger than what can be directly applied in the lab. These internal electric fields can be aligned in the lab frame by orienting the molecules with modest external electric fields.

Our measurement uses HfF$^+$ molecular ions. In an applied electric field of $\sim\SI{58}{\volt\per\centi\meter}$, the $\tDo(v=0,J=1)$ `science' state of the molecule is split into a series of doublets as shown in Fig.~\ref{fig:apparatus_states}A. In two of these doublets, highlighted in color, the molecule is fully oriented \cite{Leanhardt2011}; the upper doublet (orange) has the intramolecular axis parallel to the applied field while the lower doublet (blue) is anti-parallel. This intramolecular axis defines the direction of an effective electric field $\Eeff\approx\SI{23}{\giga\volt\per\centi\meter}$ \cite{Meyer2006,Petrov2007,Fleig2017} acting on the spin of one of the valence electrons. In the presence of a small magnetic field, the two states in a doublet correspond to the spin of this valence electron being aligned or anti-aligned with $\Eeff$. We prepare a coherent superposition of the two states and measure the energy difference using Ramsey spectroscopy. The eEDM will give a contribution to this energy $\pm2d_e\Eeff$, with opposite sign in the two doublets. We perform the measurement simultaneously on spatially overlapping clouds of ions prepared in each of the doublets. The difference between the measured energies is our science signal.

\section{Experimental Overview}\label{sec:experimental_overview}

\begin{figure*}
    \centering
        \includegraphics{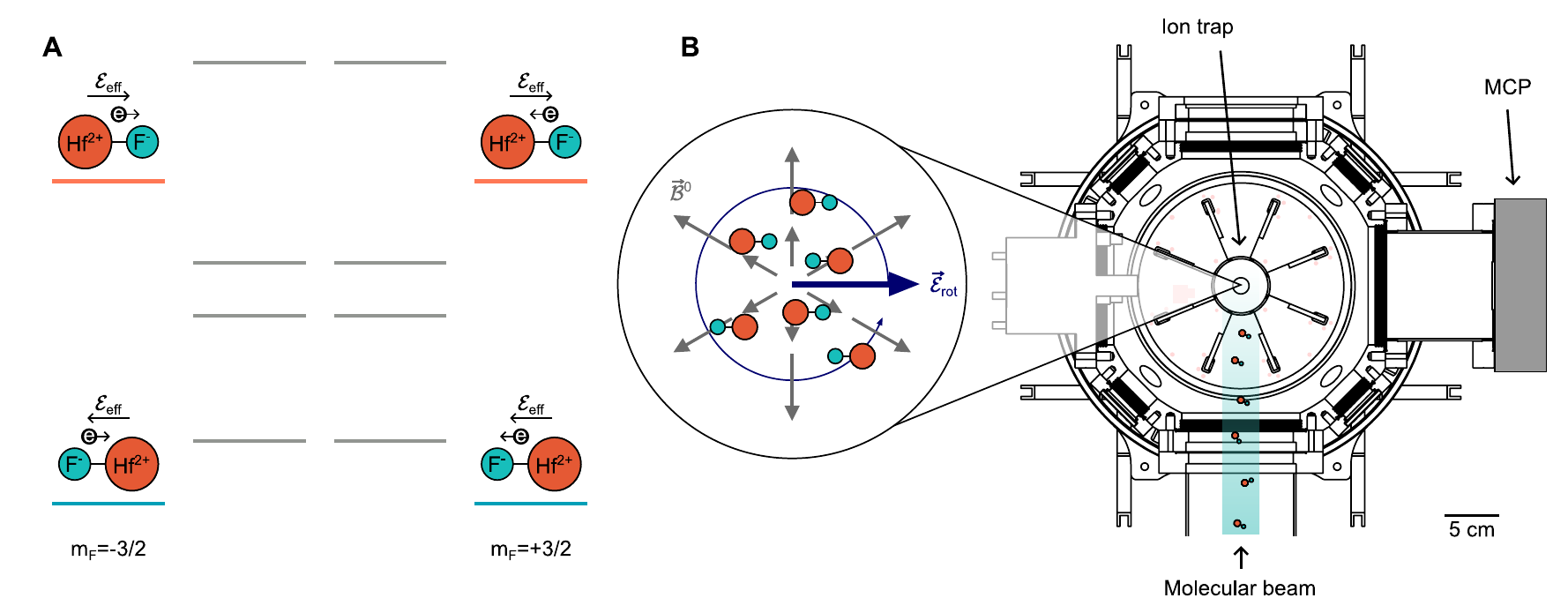}
    \caption{Experiment outline. (A) Level structure of the eEDM-sensitive $\tDo(v=0,J=1)$ state. The horizontal axis indicates $m_F$, the projection of the total angular momentum onto the externally applied electric field. The direction of the electron spin and effective electric field $\Eeff$ is indicated for each of the states used in the experiment. (B) Schematic of ion trap, composed of 8 radial electrodes and a pair of endcap electrodes. Inset shows fields applied during experimental sequence: the rotating electric bias field $\vecErot$, and the quadrupole magnetic field $\vec{\mathcal{B}^0}$.}
    \label{fig:apparatus_states}
\end{figure*}

Our experimental apparatus is shown in Fig.~\ref{fig:apparatus_states}B. An overview of the experimental sequence is given here with more details in Methods and in Ref.~\cite{Caldwell2022}. The sequence begins with production and radiofrequency trapping of roughly $20,000$ HfF$^+$ ions. In order to orient the molecules while maintaining confinement, we rotate the orienting field $\vecErot$ at angular frequency $\wrot=2\pi\times\SI{375}{\kilo\hertz}$ and perform our spectroscopy in this rotating frame. We also apply a quadrupole magnetic field gradient to create an effective bias magnetic field $\Brot$ (see Methods).

\begin{figure*}
    \centering
        \includegraphics{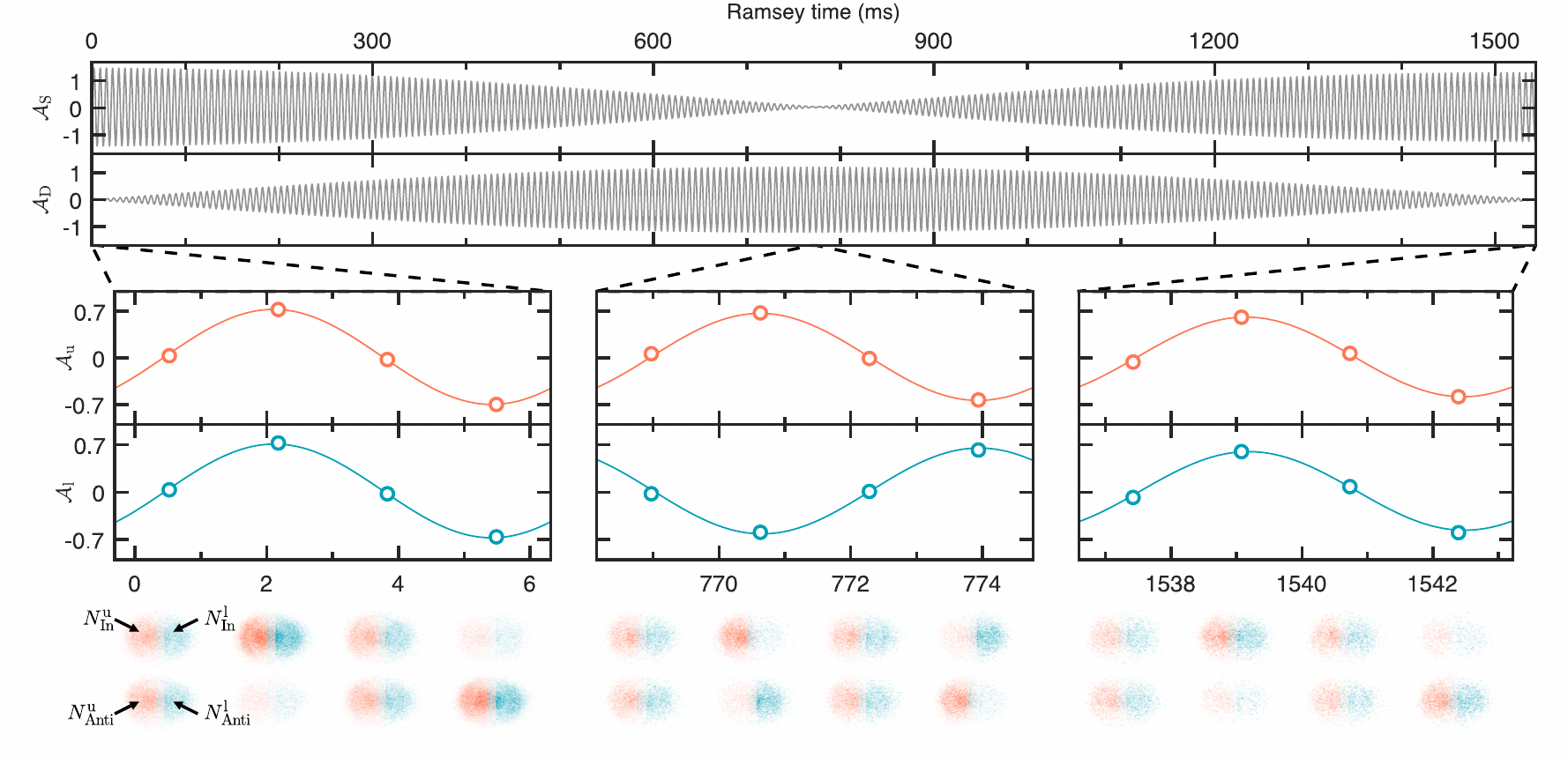}
    \caption{Example Ramsey data. Bottom two rows: detection of Hf$^+$ ions; ions are assigned to the upper or lower doublet based on their position, orange for the upper doublet, blue for the lower doublet. Counts from a thin central swatch where the assignment is ambiguous, shown in gray, are removed. Images shown are averaged over 60 shots of the experiment. Middle row: asymmetries for the upper and lower doublet. Top row: sum and difference asymmetries used to extract frequencies. Note that middle-time data, where the two doublets are out of phase, is shown only for illustrative purposes and was not collected during the dataset.}
    \label{fig:ramsey_fringe_figure}
\end{figure*}

We prepare an incoherent mixture of one of the spin states from each doublet, either $m_F=3/2$ or $m_F=-3/2$, and then apply a $\pi/2$ pulse to create a coherent superposition of the two states in each doublet. We allow the superposition to evolve for a variable amount of time, then apply a second $\pi/2$ pulse to map the accumulated relative phase onto a population difference between the two states. We clean out the population in one of the spin states in each doublet, then detect and count the number of ions in the remaining stretched state by state-selectively photodissociating the molecules \cite{Ni2014}. We use the opposing orientations of the two doublets in the trap to send the dissociation products to opposite sides of our imaging microchannel plate (MCP) and phosphor screen assembly \cite{Shagam2020,Zhou2020}. We then repeat the procedure with the opposite initial spin state. Example data are shown in the two rows at the bottom of Fig.~\ref{fig:ramsey_fringe_figure}. We count the ions on each side of the screen, in each configuration---typically $\sim120$ ions from each doublet after \SI{3}{\second} hold time---and from these two measurements construct the two asymmetries, $\mathcal{A}_{\rm u}$ and $\mathcal{A}_{\rm l}$, where
\begin{equation}
\mathcal{A}_{\rm u/l} = \frac{N^{\rm u/l}_{\rm In}-N^{\rm u/l}_{\rm Anti}}{N^{\rm u/l}_{\rm In}+N^{\rm u/l}_{\rm Anti}}.
\end{equation}
Here $N^{\rm u/l}_{\rm In}$ and $N^{\rm u/l}_{\rm Anti}$ are the number of ions counted; $\rm{u/l}$ indicate the upper or lower doublet and the subscripts indicate whether we read out the same state we prepare (In) or the opposite (Anti). We repeat our measurement at different free-evolution times, generating a pair of Ramsey fringes as shown in the middle row of Fig.~\ref{fig:ramsey_fringe_figure}. The frequencies of these two fringes are proportional to the energy splitting in the two doublets. The primary contribution to this energy splitting is the Zeeman splitting, $3 g_F \mu_{\rm B} \Brot$, where $g_F=\num{-0.0031+-0.0001}$ \cite{Loh2013} is the $g$-factor of the science state. For our typical experimental parameters, this produces fringe frequencies of $\sim\SI{100}{\hertz}$. Other effects including the eEDM make small modifications to this frequency.

Instability of the intensity of the pulsed lasers used for creation and photodissociation of the ions (see Methods) creates considerable noise in the number of ions measured in each shot of the experiment, typically $\sim3\times$ the quantum-projection-noise limit on the side of the fringe at \SI{3}{\second}. However these sources of noise, and many others, are common mode between the two doublets---the exact same laser pulses address both clouds of ions---and so the noise in $\mathcal{A}_{\rm u}$ and $\mathcal{A}_{\rm l}$ is highly correlated. To take advantage of this, we form the sum and difference asymmetries $\mathcal{A}_{\rm S}=\mathcal{A}_{\rm u}+\mathcal{A}_{\rm l}$ and $\mathcal{A}_{\rm D}=\mathcal{A}_{\rm u}-\mathcal{A}_{\rm l}$ as shown in the top row of Fig.~\ref{fig:ramsey_fringe_figure}. If we take data when the two doublets are close to in phase, the noise in $\mathcal{A}_{\rm D}$ is drastically reduced \cite{Zhou2020}. The two doublets oscillate at slightly different frequencies owing to a $\sim1/230$ difference in their magnetic moments, and so during the eEDM dataset we deliberately take our data at a \textit{beat}. We take two sets of points; the early-time data when the two doublets are in phase, and the late-time data $\sim 230$ oscillations later when they come back into phase again. We can control the time of the second beat by varying the strength of the magnetic bias field $\Brot$. We fit to $\mathcal{A}_{\rm S}$ and $\mathcal{A}_{\rm D}$ to extract the mean of the two fringe frequencies and their difference.

We collect Ramsey fringes in $2^3=8$ experimental states, corresponding to each possible combination of 3 binary experimental switches $\{\tilB,\tilR,\tilI\}=\pm1$. $\tilB$ is the direction of the magnetic bias field relative to $\vecErot$, $\tilR$ the rotation direction of $\vecErot$, and $\tilI$ the direction of $\vecErot$ relative to the imaging MCP at the instant of photo-dissociation--determining which side of the phosphor screen each of the doublets is imaged onto. A set of Ramsey fringes in each of the 8 switch states forms a block. To minimize the effects of experimental drifts, within a block we interleave data collection for the switch states; the first Ramsey time is recorded for all switch states before moving onto the second Ramsey time for each switch state etc. We take the 16 fitted frequencies from each block and form 16 linear combinations to give the components of the measured frequencies which are even or odd under each of the experimental switches. Following Ref.~\cite{ACME2014}, we label the components with superscripts that denote the switches under which the quantity is odd. Our science signal is $\fBD$, the component of the difference frequency which is odd under $\tilB$ but even under $\tilR$ and $\tilI$ \footnote{Note that we define the mean frequency to be always positive. If we had instead allowed the fringe frequency to change sign when the sign of the magnetic bias field changes, our science signal would have been $\fD$.}. The other channels allow us to diagnose systematics and monitor experimental performance. 

Over the course of the dataset, we varied a number of other experimental parameters on timescales slower than a block. These include: the state we read out at the end of the Ramsey sequence denoted $\tilP=\pm1$, alternated each block; the order in which the switch states are recorded at each Ramsey time, alternated every other block; three different magnitudes of the magnetic bias field, corresponding to mean fringe frequencies of $\fn\sim\SI{77}{\hertz}$, \SI{105}{\hertz} and \SI{151}{\hertz}; and reversing the waveplates that set the lab-frame handedness of the light used for state preparation and readout. During data collection and analysis, we `blinded' our measurement of $\fBD$ by adding an unknown offset to this channel. The offset was not removed until our systematics search and analysis \cite{Caldwell2022} were complete.


\section{Accuracy evaluation}\label{sec:accuracy_evaluation}
\begin{figure}
    \centering
        \includegraphics{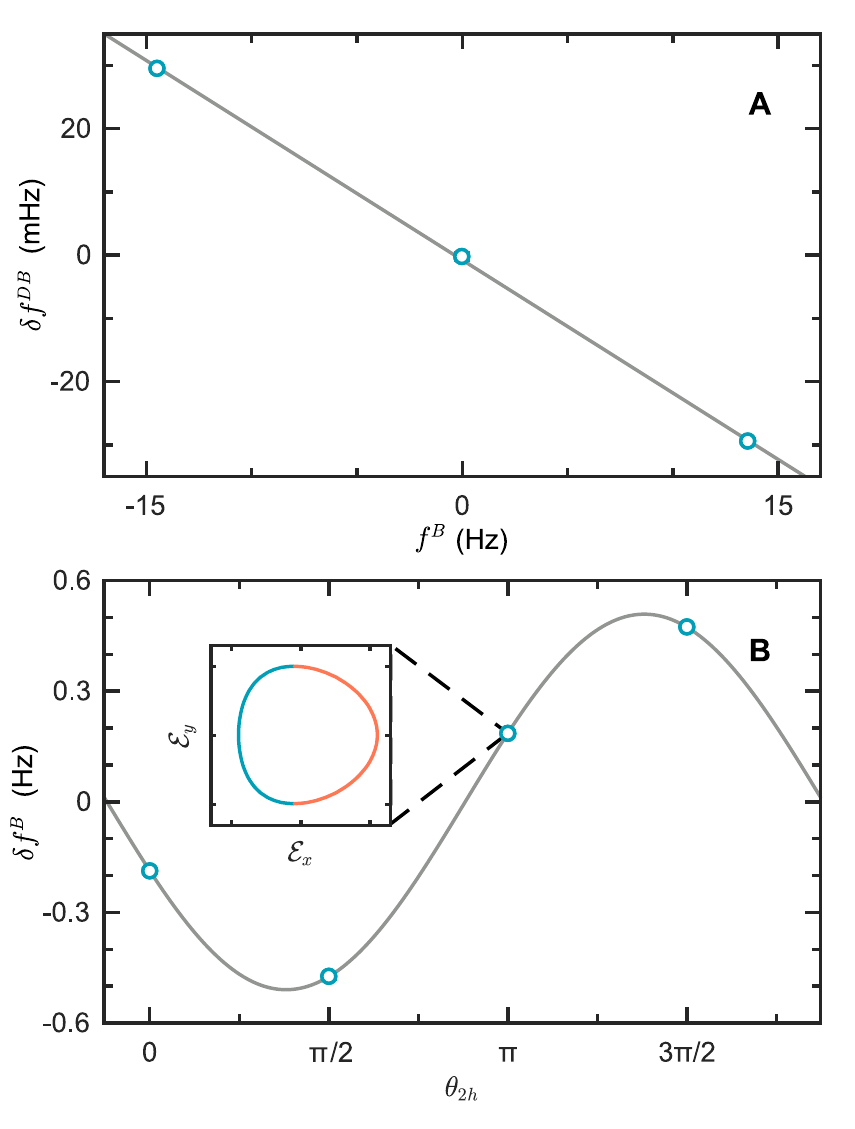}
    \caption{Systematic shifts in the measurement. (A) Shift in $\fBD$ due to non-reversing axial magnetic field, which can be corrected using the $\sim460$ times larger shift in the $\fB$ channel. (B) Shift in the $\fB$ channel due to deliberately applied second-harmonic electric field $\mathcal{E}_{2h}$ with transverse magnetic field $\mathcal{B}$. Data shows variation in shift as angle $\theta_{2h}$ between $\mathcal{E}_{2h}$ and $x$ axis is varied. $\mathcal{E}_{2h}$ is $\sim250$ times larger than that present in the experiment, $\mathcal{B}$ is $\sim\SI{14}{\milli\gauss}$. Inset shows Lissajous figure traced out by total electric field for greatly exaggerated ratio of $\mathcal{E}_{2h}/\Erot$; blue and orange show one half cycle each. The field points in the $-x$ direction for more time than in the $+x$ direction.}
    \label{fig:fB_vs_fBD}
\end{figure}

To evaluate the accuracy of our measurement, we searched extensively for systematic shifts prior to data collection; a summary is given in Table~\ref{tab:systematic_error_budget}. In general, we tuned a variety of experimental parameters over ranges large compared to those present during data collection, exaggerating any accompanying systematic effects, and observed the response in our data channels. The only shift we were able to observe directly in the eEDM channel is due to a non-reversing quadrupole magnetic field and the difference in magnetic moments between the two doublets, caused primarily by the applied electric field mixing the states of the two doublets with higher rotational levels of the molecule. The $\fB$ channel provides a direct measurement of the non-reversing magnetic field and allows us to apply a correction to our science channel, $\delta \fBD_{\rm corr} = \fB \frac{\dgF}{g_F}$, where $\dgF$ is half the difference between the $g$-factors for the upper and lower doublets, see Fig.~\ref{fig:fB_vs_fBD}A. Before applying any corrections to the science channel we suppress this systematic by actively shimming the currents through the coils which apply the magnetic bias field to minimize $\fB$. This shimming was so effective that the mean correction we applied was well below our statistical sensitivity, $\delta \fBD_{\rm corr} = \SI{90}{\nano\hertz}$. 

Unfortunately, this type of shimming and correction leaves us susceptible to other possible effects which cause shifts in $\fB$ and $\fBD$ with a ratio different from $\frac{\dgF}{g_F}$. An important example of such a shift is the combination of a transverse magnetic field with an electric field oscillating at $2\wrot$, present in our experiment due to harmonic distortion in the amplifiers driving the trap electrodes. The harmonic distortion causes the electric field, and thus the magnetic moment of the molecule, to spend more time pointing in one spatial direction than the other giving a non-zero time-averaged Zeeman interaction with a background magnetic field. This causes shifts in $\fB$ but no corresponding shift in $\fBD$ where the effect is canceled by the differential magnetic moment also changing size due to the distortion. Figure~\ref{fig:fB_vs_fBD}B shows the shift in $\fB$ when we deliberately apply a second-harmonic electric field and vary its angle. We shim out the second harmonic on each electrode by feeding forward a second-harmonic signal with the opposite phase, suppressing the amplitude by $\sim 80$\,dB. We used magnetic shim coils to null the ambient magnetic field at the trap center below \SI{10}{\milli\gauss}. The measured sizes of the residual effects were used to compute the maximum size of the systematic during our dataset.

A full account of all systematic shifts considered is presented in \textcite{Caldwell2022}.

\begin{table}
	\caption{Summary of systematic shifts and their uncertainties presented in \textcite{Caldwell2022}. All values are in \si{\micro\hertz}.}
	\centering
	\begin{ruledtabular}
	\begin{tabular}{@{}l r r@{}}
	    Effect  & Correction & Uncertainty\\
	    \midrule
	    Magnetic\\
	    \quad Non-reversing $\vec{\mathcal{B}^0}$  & 0.1 & $<0.1$  \\ 
        \quad Stray $\mathcal{B}$ fields + distortion of $\Erot$  & & 3.2  \\ 
	    Berry's phase\\
        \quad Rotation-odd axial secular motion &  & 3.4  \\ 
        \quad Axial fields at harmonics of $\Erot$ &  & 3.4  \\ 
        Simultaneous doublet spectroscopy\\
        \quad Imperfect spatial overlap  & & 3.5  \\
        \quad Imperfect imaging contrast &  & 1.4  \\
	    Other\\
	    \quad Rotation-induced $m_F$-level mixing  & & 0.4 \\
        \addlinespace
        Total &  0.1 & 6.9
	\end{tabular}
	\label{tab:systematic_error_budget}
	\end{ruledtabular}
\end{table}

\section{Results}\label{sec:results}

\begin{figure}
    \centering
        \includegraphics{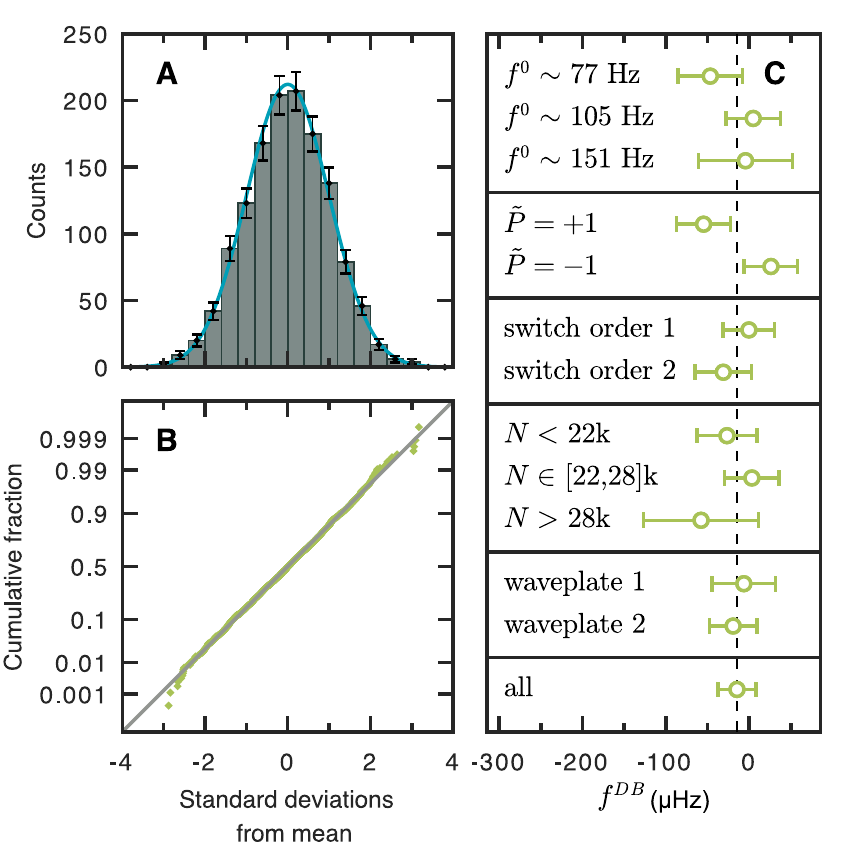}
    \caption{Summary of our dataset after cuts have been applied and the uncertainty on each block scaled by $\sqrt{\chi^2}$ to account for over-scatter. (A) Histogram of data. Error bars show standard deviation of bin counts expected from Poisson distribution. Blue line shows normal distribution. (B) Normal probability plot of $\fBD$ showing the data are consistent with a normal distribution. Gray line shows expected probability for a normal distribution. (C) Variation of central value under different experimental parameters compared to the overall average value of $\fBD$. Here $N$ is the average number of trapped \hffp{} ions per experimental trial during a block. Other panels are described in the final paragraph of Sec.~\ref{sec:experimental_overview}.}
    \label{fig:statistics_figure}
\end{figure}

We collected 1370 blocks over roughly two months, corresponding to $\sim 620$ hours of data and $\sim 10^8$ ion detection events. Each block results in one value of $\fBD$ and thus a single measurement of $d_e$. The uncertainty on $\fBD$ for each block is calculated using only the standard errors on the asymmetries for that block. We applied cuts to the blinded data based on non-eEDM channels indicating signal quality. Blocks with late-time contrast below 0.2 were cut due to low signal-to-noise, as were blocks containing fitted fringe frequencies which were over $3.5\sigma$ different from the mean fringe frequency for that switch state. After applying cuts we were left with 1329 blocks with $\chi^2=\num{1.07+-0.04}$ for $\fBD$. Figure~\ref{fig:statistics_figure}A and B show the distribution of measured $\fBD$ values over the 1329 blocks after relaxing the uncertainty for each of the blocks by a factor $\sqrt{\chi^2}=1.035$. The data are consistent with a normal distribution. Our final statistical uncertainty of \SI{22.8}{\micro\hertz} is obtained using these relaxed uncertainties. Based on the number of ions detected in each shot, this uncertainty is $\sim30\%$ above the quantum-projection-noise limit.

Figure~\ref{fig:statistics_figure}C shows how the measured value of $\fBD$ depends on experimental parameters varied during the dataset; we find no concerning dependencies.

We removed our blind on November 1, 2022 and obtained a final value for the eEDM-sensitive frequency channel
\begin{equation}
    f^{BD} = \SI[parse-numbers=false]{-14.6 \pm 22.8 _{\rm stat} \pm 6.9 _{\rm syst}}{\micro\hertz}.
\end{equation}
Dividing by $-2\Eeff  \mathrm{sgn}(g_F)/h \simeq \SI{1.11e31}{\micro\hertz\per\electron\per\centi\meter}$ \cite{Fleig2018}, we obtain a value for the eEDM
\begin{equation}
    d_e = \SI[parse-numbers=false]{(-1.3 \pm 2.0_{\rm stat} \pm 0.6_{\rm syst})\times 10^{-30}}{\electron\centi\meter},
\end{equation}
consistent with zero within one standard error. The combined statistical and systematic uncertainty $\sigma_{d_e}=\SI{2.1e-30}{\electron\centi\meter}$ improves on our previous work \cite{Cairncross2017} by a factor $\sim37$, and on the previous state-of-the-art from the ACME collaboration \cite{ACME2018} by a factor $\sim2$. Our new result and those of the ACME collaboration---two measurements using radically different experimental platforms with contrasting sources of systematic shifts---are consistent at slightly above one standard error.

\section{Interpretation}

\begin{figure}
    \centering
    \includegraphics{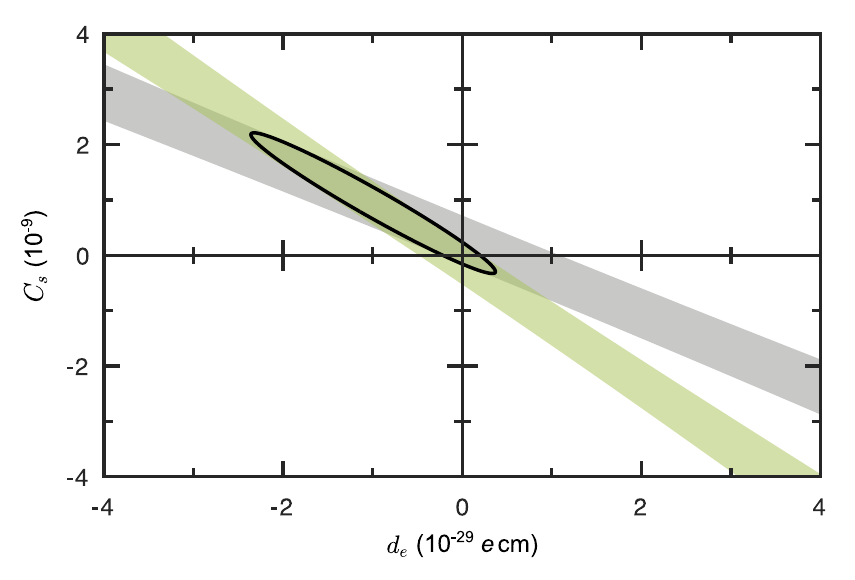}
    \caption{Fit to results of this work and Ref.~\cite{ACME2018}. Green and grey shaded regions show 90\% confidence bands for HfF$^+$ and ThO respectively. Ellipse shows 90\% confidence limit for global fit. Parameters used in fits are from Ref.~\cite{Fleig2018}.}
    \label{fig:global-analysis}
\end{figure}

We use our result to obtain an upper bound using a folded gaussian distribution,
\begin{equation}\label{eq:limit}
    | d_e | < \SI{4.1 e-30}{\electron\centi\meter}\;\rm{(90\%\;confidence)}.
\end{equation}
This limit constrains extensions to the Standard Model which predict new sources of CP-symmetry violation to explain the matter--anti-matter asymmetry of the universe \cite{Canetti2012}.  Many extensions, including supersymmetry, the two-Higgs model, and left-right symmetric models, generate an eEDM at the one-loop level \cite{Chupp2019}, with magnitude \cite{Engel2013}
\begin{equation}
    d_e \sim \dfrac{e a_0 \alpha}{2} \dfrac{g^2}{2\pi} \sin{\phi_{CP}}\dfrac{m_e^2}{M^2}.
\end{equation}
Here $M$ is the characteristic mass of new particles with effective coupling strength $g$ to the electron, and $\phi_{CP}$ is the phase which describes how strongly the interaction violates CP symmetry.  Because $d_e \propto M^{-2}$ and our limit in Eq.~\ref{eq:limit} is $\sim2.4$ times smaller than the one reported in \cite{ACME2018}, we are sensitive to new particles with $\sqrt{2.4} = 1.5$ times higher mass.

To estimate the mass reach of our experiment, we need to make assumptions for the size of $g^2$ and $\sin{\phi_{CP}}$. For the strong force, QED and the weak force, $g^2 \approx 1, 1/137$ and $10^{-6}$ respectively. For extensions to the Standard Model seeking to explain the matter--anti-matter asymmetry, we expect $\sin{\phi_{CP}} \sim 1$. With this assumption, we can interpret our new limit on $d_e$ as $M/g \gtrsim \SI{40}{\tera\eV}/\alpha^{1/2}$, an order of magnitude beyond the direct reach of the Large Hadron Collider \cite{Bose2022}.

So far we have assumed that CP violation arises purely from $d_e$. Diatomic molecules are also sensitive to pseudoscalar-scalar electron-nucleon coupling $C_S$ \cite{Chupp2015,Denis2015}, and we can interpret our measurement as a linear combination $h \fBD\times\mathrm{sgn}(g_F)=-2 \Eeff d_e+2 W_S C_S$ where $W_S/h=\SI{-51}{\kilo\hertz}$ \cite{Fleig2018} is a molecule-specific structure constant. Assuming that $d_e$ is zero, we can instead attribute our measurement to $C_S$, we find
\begin{equation}
    C_S = (-1.4 \pm 2.2 _{\rm stat} \pm 0.7 _{\rm syst})\times 10^{-10}.
\end{equation}

Determining rigorous limits on $d_e$ and $C_s$ requires combining the results of two or more measurements using molecules with different ratios of $\Eeff$ to $W_S$. Figure~\ref{fig:global-analysis} shows a combined fit to the results of this work and Ref.~\cite{ACME2018}, giving upper bounds of $|d_e|<\SI{2.1e-29}{\electron\centi\meter}$ and $|C_S|<\num{1.9e-9}$ with 90\% confidence. Our measurement improves these bounds by factors of 16 and 12 respectively \footnote{Improvements are given with respect to equivalent bounds calculated from combined fit to results of Ref.~\cite{ACME2018} and Ref.~\cite{Cairncross2017}.}. 

\begin{acknowledgments}
We thank the staff at JILA for making this experiment possible. 
This work was supported by the Marsico foundation, the Sloan Foundation, the Gordon and Betty Moore Foundation, NIST, and the NSF (Award PHY-1125844). TW acknowledges funding support from NSF GRFP.

\end{acknowledgments}


\bibliography{references}

\appendix
\setcounter{figure}{0}
\makeatletter 
\renewcommand{\thefigure}{S\@arabic\c@figure}
\makeatother

\section*{Materials and Methods}
\label{sec:methods}

\subsection{Apparatus}

The experiment begins with a pulsed beam of neutral HfF molecules, produced by ablation of a solid Hf rod into a supersonic expansion of an Ar and SF$_6$ gas mixture. The Hf plasma reacts with the SF$_6$ to create HfF which is entrained in the Ar flow and cooled by collisions with the Ar atoms. When the pulse of molecules reaches the main chamber (shown in Fig.~\ref{fig:apparatus_states}B) $\sim\SI{50}{\centi\meter}$ away, a pair of pulsed UV lasers at \SI{309}{\nano\meter} and \SI{368}{\nano\meter} state-selectively ionize the molecules \cite{Loh2011,Loh2013}. The newly created ions are confined in an RF Paul trap composed of eight radial electrodes and two endcap electrodes. We drive the radial electrodes in pairs at \SI{50}{\kilo\hertz} to produce the oscillating quadrupole field which provides radial confinement (plane of the page in Fig.~\ref{fig:apparatus_states}B) and apply DC voltages to the two endcaps to provide axial confinement (perpendicular to plane of page in Fig.~\ref{fig:apparatus_states}B). During the Ramsey spectroscopy, the trapping frequencies are $\sim\SI{1}{\kilo\hertz}$ in each direction. 

We apply the orienting field $\vecErot=\Erot \left[ \hat{x} \cos{(\wrot t)} + \tilR \hat{y} \sin{(\wrot t)} \right]$ by driving the radial electrodes at $\wrot=2\pi\times\SI{375}{\kilo\hertz}$, with each electrode $\pi/4$ out of phase with its neighbor. The direction of rotation $\tilR=\pm1$ is set by switching the sign of the sign of the relative phase. During data collection $\Erot\sim\SI{58}{\volt\per\centi\meter}$. The bias field induces additional circular micromotion of the trapped ions $\vec{\delta r}=-\rrot\hatErot$, with $\rrot\sim\SI{0.5}{\milli\meter}$. The magnetic bias field is realized by using a pair of coils in anti-Helmholtz configuration, aligned with the axial direction to generate a quadrupole magnetic field gradient, $\vec{\mathcal{B}^0}=\tilB \mathcal{B}_{2,0}^{\rm rev} (2 \vec{z} - \vec{x} - \vec{y})$, where $\tilB=\pm1$. In the presence of the $\vecErot$-induced micromotion, this produces an effective bias field $\Brot=\tilB\mathcal{B}_{2,0}^{\rm rev}\rrot$.

We infer the magnetic field at the position of the ions using measurements from an array of eight 3-axis fluxgate magnetometers bolted to the exterior of the main chamber. The magnetic field can be controlled using 3 pairs of shim coils aligned with the lab $x$, $y$ and $z$ directions.

\subsection{State preparation}

\begin{figure}
    \centering
        \includegraphics{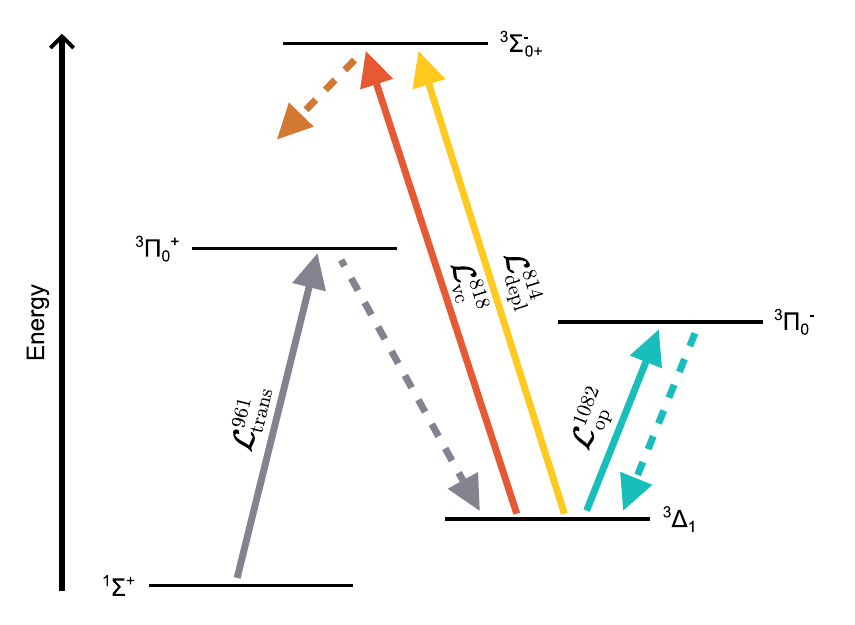}
    \caption{Cartoon depicting the transitions used during our state preparation and labels for the lasers used to address them. The ${}^3\Pi_{0^+}$ and ${}^3\Pi_{0^-}$ states decay preferentially to ${}^3\Delta_1$, while the ${}^3\Sigma_0^-$ state decays preferentially to ${}^1\Sigma^+$.}
    \label{fig:state_prep_level_diagram}
\end{figure}

State preparation is facilitated by four CW lasers depicted in Fig.~\ref{fig:state_prep_level_diagram}. The state-selective two-photon ionization procedure produces \hffp in the ground electronic and vibrational state $\sSp(v=0)$, primarily distributed over the lowest 4 rotational levels. We connect these rotational levels using microwaves and perform incoherent transfer to the $\tDo(v=0,J=1)$ science state (see Fig.~\ref{fig:apparatus_states}A) by applying \ltrans{} light at \SI{961}{\nano\meter} addressing the $\tPz(J=0) \leftarrow \sSp$ transition. The decay from $\tPz$ populates several vibrational levels in $\tDo$, which decay into our science state during free evolution if left untreated. To avoid this, we use \lvc{} light at $\SI{818}{\nano\meter}$ which connects $\tSm(v=1, J=0) \leftarrow  \tDo, (v=1, J=1)$, where $\tSm$ preferentially decays back to $\sSp$. This laser is on for the duration of the experiment, and cleans out any population which leaks down from higher vibrational levels over the course of the measurement. 

After transfer to $\tDo$ we turn on \lop{} at $\SI{1082}{\nano\meter}$ to drive $\tPzm(v=0, J=0) \leftarrow  \tDo(v=0, J=1)$. The light is circularly polarized with its $k$-vector in the plane of $\vecErot$, and strobed synchronously with it such that it drives a $\sigma^{+/-}$ transition to an $F' = 3/2$ manifold in $\tPzm$. This optically pumps population to an incoherent mixture of the spin states in each of the two doublets used for the measurement. Any residual population in other states is cleaned out using \ldepl{} light at \SI{814}{\nano\meter} to drive the $\tSm(v=0, J=0) \leftarrow \tDo(v=0, J=1)$ transition, which decays preferentially to $\sSp$. This light is also circularly polarized, with the same $k$-vector and handedness as \lop{}. The molecules are now ready for the Ramsey sequence.

\subsection{Ramsey spectroscopy}

The rotation of $\Erot$ induces a mixing between the two states in a doublet \cite{Meyer2009,Leanhardt2011}. The mixing is proportional to $1/\Erot^3$ and is very weak at $\Erot\sim\SI{58}{\volt\per\centi\meter}$. We apply our $\pi/2$ pulses by ramping down the magnitude of $\Erot$ to $\sim1/12$ of its initial value, dramatically increasing the mixing and causing the pure spin state to evolve into a coherent superposition. We hold the lower value of $\Erot$ for \SI{1}{\milli\second} before ramping back up and allowing the superposition to evolve for the variable Ramsey time. We then repeat the process to map the accumulated phase onto a population difference between the two states in a doublet.

\subsection{Measurement}

We project the ions into their final state by reapplying \ldepl{} light to clean out population in one of the spin states in each doublet. The timing of the strobing is chosen to apply light of either the same or opposite handedness to that used in state preparation, depending on whether we are performing the `In' or `Anti' chop. We detect and count the number of ions in each of the remaining spin states by photodissociating them and then ejecting them onto an microchannel plate (MCP) and phosphor screen assembly. The photodissociation is driven by two pulsed UV lasers at \SI{368}{\nano\meter} and \SI{266}{\nano\meter}. The resultant \hfp{} ions from each doublet receive kicks in opposite directions corresponding to the orientation of the doublets in the trap at the instant of dissociation. The dissociation pulse is timed so that $\vecErot$ is in the plane of the MCP. The \hfp{} ions from each doublet are resolved from residual \hffp{} ions by time of flight, and from each other by imaging which side of the phosphor screen they impact. Technical details of our imaging and counting system are given in Refs.~\cite{Shagam2020} and \cite{Zhou2020}. 

\subsection{Improvements on Generation I}

The uncertainty on our measurement is improved by a factor $\sim37$ relative to the first generation of this experiment \cite{Cairncross2017}, facilitated by a redesigned ion trap and improvements to state preparation and readout. The increased volume of the new trap enables more ions to be confined at the same density and, combined with improved state-preparation efficiency realized by incorporating optical pumping techniques, results in an increase in the number of ions detected in each shot by a factor $\sim40$. The new trap also provides a larger, more uniform, rotating electric field, extending the coherence time of the ions by a factor $\sim4$. Finally, we developed techniques to simultaneously detect the upper and lower doublets in a single shot of the experiment \cite{Zhou2020}, enabling cancellation of the largest sources of technical noise and rejection of many important sources of systematic error.

\subsection{Interpretation}

Our measurement provides a best estimate on $d_e$, $\mu_{d_e}=\SI{-1.3e-30}{\electron\centi\meter}$ and a combined statistical and systematic uncertainty $\sigma_{d_e}=\SI{2.1e-30}{\electron\centi\meter}$. To calculate an upper bound on $|d_e|$ with confidence $P$, we find the value $L$ which satisfies
\begin{equation}
    \int_{-L}^L \frac{1}{\sqrt{2\pi}\sigma_{d_e}}e^{-\frac{(x-\mu_{d_e})^2}{2\sigma_{d_e}^2}} = P.
\end{equation}
We note that this is different from Feldman-Cousins prescription used by the ACME collaboration \cite{ACME2017,Feldman1998}; our 90\% confidence interval calculated using this prescription is $|d_e| < \SI{4.5e-30}{\electron\centi\meter}$. When using either method to calculate the limits from both experiments, the improvement in the limit is $\sim2.4$.


\end{document}